**Correspondence on "ACMG STATEMENT: ACMG SF v3.0 list for reporting of secondary findings in clinical exome and genome sequencing: a policy statement of the American College of Medical Genetics and Genomics (ACMG)" by Miller et al.**

We were interested to read the recent update on recommendations for reporting of secondary findings in clinical sequencing[1], and the accompanying updated list of genes in which secondary findings should be sought (ACMG SF v3.0)[2]. Though the authors discuss challenges around incomplete penetrance in considerable detail, we are concerned that the recommendations do not fully convey the degree of uncertainty regarding the penetrance of variants in genes associated with inherited cardiomyopathies, which make up almost a quarter of the list. Since penetrance is incomplete and age-related, individuals found to carry variants will often require surveillance, rather than a one-off definitive diagnostic assessment. There is a lack of evidence regarding benefits, harms, and healthcare costs associated with opportunistic screening.

Here, we review the data from the studies cited to support the inclusion of two new dilated cardiomyopathy (DCM)-associated genes, *FLNC* and *TTN*, alongside other published data, and provide new analyses of publicly available data. Many of our conclusions may also be applicable to genes included in the previous ACMG SF v2.0. Of note, the ACMG/AMP standards have been calibrated for variants found in people with confirmed disease: we do not discuss here the further challenges in identifying which variants have disease-causing potential outside this context.

There are many challenges in assessing and reporting penetrance. Many penetrance estimates come from studies in families of affected individuals, where penetrance may be higher than in the wider population. Individuals found to carry (likely) pathogenic variants (P/LP) in genotype-first analyses can be considered in four groups: (a) known affected, (b) undiagnosed affected, (c) unaffected but will develop disease, (d) will never develop disease. Group (a) are, by definition, outside the scope of these recommendations but they are often included in studies estimating penetrance, which is appropriate for some questions, but may over-estimate the benefits of opportunistic screening. Cohorts used in genotype-first analyses may also be enriched or depleted for this group according to ascertainment approach. One-off assessment will detect (b) but not (c), for whom burdensome and costly longitudinal surveillance may be required.

*FLNC*

"*The SFWG voted to include this gene based on its high penetrance, severity of the phenotype if untreated, and the strong potential benefit of intervention based on returning P/LP variants in this gene as a SF*"[2]. While we agree with the comments on apparent phenotypic severity in *FLNC*-related DCM, we are not aware of any data from population studies to justify an assertion of high penetrance outside of families with known disease. The recommendations cite a family-based analysis[3], a cardiomyopathy case series[4], and a review article[5]. The review authors note that *"the finding of a truncating FLNC variant in otherwise healthy subjects outside of a familial context is much less clear at the moment, as there is not enough knowledge regarding penetrance, expression, and clinical correlation"*[5].

We therefore performed an analysis in 200,581 UK Biobank (UKBB) participants with exome sequencing data available (median age 58 at recruitment). We identified 50 individuals heterozygous for 38 rare truncating variants in *FLNC* (FLNCtv; prevalence 0.025%) that would be considered P/LP in an individual with DCM. The prevalence is comparable to gnomAD (47 heterozygous individuals in 125,408 = 0.037%). Among these 50 participants, there were no cases of DCM, or other inherited cardiomyopathy, and DCM was not identified in the five individuals with cardiac MRI, which we have previously found to have higher sensitivity than ICD codes alone[6]. Lifetime risk of major adverse cardiovascular events (MACE, see Supplementary Material) was higher in *FLNC* variant heterozygotes (HR=1.9, P=0.04), driven by increased risk of atrial fibrillation and arrhythmia (HR=2.4,

P=0.0096), but with modest absolute increase. There were five deaths, three heart failure events, and no cardiac arrests, in 569 person-years of follow-up, which was not significantly different from the rest of the population (Table S2).

*TTN*

The authors of the recommendations found that "*new evidence indicated significant risk for cardiomyopathy among those with TTN truncating variants (*TTNtv*)*"[2], citing a study of two cohorts drawn from health systems[7]. The prevalence of DCM in the cohorts was higher than population estimates, consistent with ascertainment on the basis of disease, as might be expected in a health system, and as reported by the study authors[7]. The proportion of TTNtv+ individuals who manifested DCM in these cohorts was 30% and 7.5%, which is likely an over-estimate of penetrance in an unselected population. Incident cases were not reported.

In UKBB we identified 877 participants (0.44%) carrying one or more of 487 rare TTNtv that would be reportable, similar to previous estimates[8]. We estimated the prevalence of known cardiomyopathy in TTNtv heterozygotes as 1.4% at enrolment (excluding coronary disease and HCM; Table S2). These participants with known disease may benefit from a molecular diagnosis reported as a secondary finding if not already tested.

Amongst those TTNtv heterozygotes not coded with cardiomyopathy who underwent cardiac MRI, 2.4% met criteria for DCM. This estimates the yield of a one-off cardiac assessment following reported secondary findings.

A further 3.4% TTNtv heterozygotes developed cardiomyopathy subsequently to the 1.4% at enrolment (Table S2), yielding ~3 incident cases per 1,000 person-years of surveillance (Figure S1), consistent with previous reports[8]. This estimates the yield of ongoing surveillance in those not diagnosed at first assessment, the costs and harms of which have not been well characterised to our knowledge. We observe an increased lifetime risk of MACE in TTNtv heterozygotes (HR=2.6, P=<0.001), driven by increased risk of atrial fibrillation and arrhythmia (HR=2.7, P=<0.001), HF (HR=4.4, P=<0.001), and CM (HR=15.0, P=<0.001), albeit with a small absolute increase and no significant difference in death, cardiac arrest, or stroke, with a total of 141 MACE during 10,132 person-years follow-up.

Estimating mortality in TTNtv-associated DCM as ~4% over 4 years[9,10], and modelling this as entirely preventable with diagnosis and treatment, we could estimate ~8,000 person-years of surveillance (1,600 CMR scans if 5-yearly imaging) would yield 25 new diagnoses of DCM, with an opportunity to prevent 1 death over the subsequent 4 years (Figure S2).

Alternatively, if we estimate the total excess mortality in TTNtv heterozygotes as 1% (over 10 years) and assume this would be fully preventable by return of secondary findings followed by long-term surveillance, then we would need to enrol 100 people into long-term surveillance to prevent one death (Supplemental Methods), even in this over-optimistic scenario.

We acknowledge the likelihood of survivorship bias in the UKBB that may skew lifelong penetrance estimates. However, the prevalence of DCM is close to population estimates (Table S1), which speaks against a substantial depletion of cases. UKBB is likely to provide reasonable estimates for opportunistic screening carried out in adults – e.g., in those undergoing sequencing for adult-onset breast or other cancers – since opportunistic screening is performed in those who did not manifest the screened-for disease earlier in life. Furthermore, an important proportion of individuals undergoing clinical sequencing will be healthy adult parents of children with rare diseases, sequenced for trio analyses.

The primary diagnoses in those undergoing clinical sequencing may also carry an adverse prognosis with competing risks that further reduce the benefits of cardiac screening and surveillance.

In summary, there is much uncertainty regarding the penetrance of variants in *FLNC* and *TTN* that can cause DCM. We do not believe that an assertion of high penetrance is justified for *FLNC*. The authors of the new recommendations acknowledge that TTNtv have low penetrance, but we provide further data to illustrate the yield of surveillance in individuals not known to have disease at first assessment. We think it is premature to recommend *TTN* screening, and thereby make this the standard-of-care, given that the costs, harms, and benefits are not yet well characterised.

**Authors and affiliations**

Kathryn A. McGurk[1], Sean L. Zheng[1,2], Albert Henry[3], Katherine Josephs[1], Matthew Edwards[2], Antonio de Marvao[4], Nicola Whiffin[5], Angharad Roberts[1,2], Thomas R. Lumbers[3,6], Declan P. O'Regan[4], James S. Ware[1,2,4,*]

[1]National Heart and Lung Institute, Imperial College London, London, UK.

[2]Royal Brompton & Harefield Hospitals, Guy's and St. Thomas' NHS Foundation Trust, London, UK.

[3]British Heart Foundation Research Accelerator, and Institute of Health Informatics, University College London, London, UK.

[4]MRC London Institute of Medical Sciences, Imperial College London, London, UK.

[5]Wellcome Centre for Human Genetics, University of Oxford, Oxford, UK.

[6]Bart's Heart Centre, St. Bartholomew's Hospital, London, UK.

*Corresponding author

Correspondence to James S. Ware - E: j.ware@imperial.ac.uk

**COMPETING INTERESTS**

J.S.W. has consulted for MyoKardia, Inc. and Foresite Labs. D.P.O. has consulted for Bayer.

**ADDITIONAL INFORMATION**

**Supplementary information** online

**Correspondence** and requests for materials should be addressed to JSW.

**Funding** This work was supported by the Wellcome Trust [107469/Z/15/Z; 200990/A/16/Z], Medical Research Council (UK), British Heart Foundation [RG/19/6/34387, RE/18/4/34215], and the NIHR Imperial College Biomedical Research Centre. The views expressed in this work are those of the authors and not necessarily those of the funders. This work was funded in part by the Wellcome Trust. For the purpose of open access, the author has applied a CC BY public copyright license to any Author Accepted Manuscript version arising from this submission.

**Data availability**

The UK Biobank biomedical database can be accessed globally for public health research. The data supporting the findings of this Correspondence are presented in the Supplementary information.

**CRediT statement**

Conceptualization: J.S.W; Data curation: K.A.M., S.L.Z., A.H., K.J., M.E., A.D.; Formal Analysis: K.A.M., S.L.Z.; Supervision: J.S.W., D.P.O., T.R.L., A.R.; Writing – original draft: K.A.M., J.S.W.; Writing – review & editing: S.L.Z., A.H., K.J., M.E., A.D., N.W., A.R., T.R.L., D.P.O.

**Ethics declaration**

The UK Biobank study (https://www.ukbiobank.ac.uk; PMID 25826379) was reviewed by the National Research Ethics Service (11/NW/0382, 21/NW/0157). Written informed consent was required and the study adheres to the principles set out in the Declaration of Helsinki. The data was de-identified. This study was conducted under terms of access approval number 47602.

**Correspondence on "ACMG STATEMENT: ACMG SF v3.0 list for reporting of secondary findings in clinical exome and genome sequencing: a policy statement of the American College of Medical Genetics and Genomics (ACMG)" by Miller et al.**

**Supplementary Information**

**Methods**

*The UK Biobank cohort*

The UK Biobank (UKBB) recruited 500,000 participants aged 40–69 years across the UK between 2006 and 2010 (National Research Ethics Service, 11/NW/0382, 21/NW/0157). Written informed consent was provided. This study was conducted under terms of access approval number 47602. 200,581 UKB participants underwent exome sequencing[1] and 21,129 of those had cardiac MRI imaging available for analyses.

*Exome sequencing data analysis and variant curation*

Variants within 100 bp of coding regions of the *TTN* and *FLNC* genes were extracted from the exome sequencing data. We identified truncating variants (tv) predicted to introduce a premature termination codon (often described as loss-of-function (LoF), i.e., frameshift, essential splice, and stop gained variants) that had a MAF of <0.1% in gnomAD and UKBB and a max FAF <0.0084%[2] in gnomAD. TTNtv were annotated with the cardiac expression of the exon impacted and variants affecting exons constitutively expressed in the heart (PSI>90%[3]) were retained. The exome sequencing data was annotated using Ensembl Variant Effect Predictor (VEP; version 104) with a plugin for gnomAD (version r2.1) and LOFTEE, and the data was organised using PLINK (version 1.90p 64-bit). The VEP output was analysed using R (version 3.6.0) and Rstudio (version 1.3.1073). Variants that were predicted by LOFTEE to be low confidence LoF variants were excluded from analyses.

*FLNC variants and DCM*

*FLNC* variants have been associated with both HCM and DCM, alongside other disorders. FLNCtv have been associated with DCM. While some missense variants are associated with HCM (especially in the ROD2 domain), there is only one missense variant observed in the UKBB that has been previously classified as P/LP for DCM, carried by eight UKBB participants (c.318C>G)[4]. This variant was originally reported in a compound heterozygous state[5] and there is not compelling evidence that this variant in the heterozygous state would be P/LP for DCM (ClinVar ID 267289). Furthermore, no other protein altering variants have evidence for DCM causation in ClinVar. Our analyses therefore focused on FLNCtv.

While FLNCtv are associated with DCM, with data suggesting an adverse clinical course, and FLNCtv appear penetrant when ascertained in families with DCM, penetrance has not been characterised in individuals not selected for DCM or a family history of disease.

*UKBB codes and datasets for phenotype analysis*

First occurrence data:

The UKBB provided first occurrence of health outcomes defined by 3-character ICD10 code using Hospital in-patient records, Death records, and Primary care records. The 'first occurrence' fields define each health outcome by the 3-character codes within ICD10's diagnostic chapters and provides the earliest date that the 3-character ICD10 (or mapped codes) was recorded through either self-report at any assessment centre, inpatient hospital data, primary care or death record data. Hospital inpatient data does not record the diagnosis date directly, but rather information about the dates the hospital episode started and ended and the dates the hospital admission started and ended. The episode start date has been used as the best proxy for the diagnosis date. If this was missing, the admission date, episode end date or discharge date were used instead. Date of the event was recorded in primary care data.

Self-report gives the interpolated year when non-cancer illness first diagnosed, obtained during verbal interview at the UKBB assessment centre.

The fields analysed were as follows:

- Any cardiomyopathy (CM): I42* excluding reported hypertrophic CM (HCM) or coronary artery disease (CAD) at 2021 data release (below) - an inclusive group of non-ischaemic & non-hypertrophic cardiomyopathic phenotypes, intended to maximise sensitivity to detect individuals with phenotypes potentially attributable to variants in *FLNC* or *TTN*
- Cardiac arrest: I46*.
- Atrial fibrillation and flutter/arrhythmia: I48* and I49*.
- Stroke: I64*.
- Heart Failure: I50*.

Date of death:

The UK Biobank receives death notifications on a regular basis through linkage to national death registries and provides the date of death.

2021 data release of refreshed Hospital Episode Statistics (HES) plus Self-Reported data:

The following codes were used to identify patients of the following conditions, with broad conditions matching the first reported data above:

- Dilated cardiomyopathy (DCM): ICD10 I420, I426, I427, O903; excluding HCM and coronary disease.
- Hypertrophic cardiomyopathy (HCM) [for exclusions]: Self-reported 1588; ICD10 I422, I421; ICD9 4251.
- Coronary disease [for exclusions]: Self-reported 1075, 1070, 1095, 1523; ICD10 I210, I211, I212, I213, I214, I219, I220, I221, I228, I229, I252; ICD9 410, 411, 412; Procedure K483, K491, K492, K493, K494, K498, K499, K751, K752, K753, K754, K758, K759, K401, K402, K403, K404, K408, K409, K411, K412, K413, K414, K418, K419, K421, K422, K423, K424, K428, K429, K431, K432, K433, K434, K438, K439, K441, K442, K448, K449, K451, K452, K453, K454, K455, K456, K458, K459.
- Coronary artery disease (CAD) [for Non-ischaemic cardiomyopathy definition]: ICD10 I21*, I22*, I23*, I23*, I241, I252, I255, I256; Procedure K40*, K41*, K42*, K43*, K44*, K45*, K46*, K47*, K48*, K49*, K50*
- Structural heart disease (HD) [for Non-ischaemic cardiomyopathy definition]: ICD10 I05*, I06*, I07*, Q20*, Q21*, Q22*, Q23*, Q24*, Q25*, Q26*; Procedure K25*, K26*, K27*, K28*, K29*, K30*, K31*, K32*, K34*, K36*, K37*, K38*
- Left ventricular systolic dysfunction (LVSD) [for Non-ischaemic cardiomyopathy definition]: ICD10 I420, I426, I427, I255, I502; Procedure K617

Non-ischaemic cardiomyopathy:

To obtain an upper estimate of variant penetrance, we used an inclusive definition of non-ischaemic cardiomyopathy (including hypokinetic non-dilated CM) defined as HF with left ventricular systolic dysfunction (LVSD) and without CAD or structural heart disease (HD). Phenotyping was performed using a rule-based algorithm based on ICD-10 and procedure codes in UKBB linked HES data and left ventricular ejection fraction (LVEF) derived from Cardiac Magnetic Resonance imaging (CMR) was used in the calculation as binary classifier for LVSD[6]. Participants were identified as cases if they did not have CAD or structural HD and had LVSD. Participants were excluded from the analysis if they had CAD or structural HD, or if they had: 1) HF without history of LVSD or 2) CAD or structural HD. LVSD was defined as LVEF <50 or the presence of LVSD codes.

Analysis of cardiac MRI data:

Summary CMR data was analysed as previously described[6]. Participants were flagged for DCM if they had increased LV end-diastolic volume (LVEDV) (women >175mL and men >232mL[7] with LVEF <50% and no reported HCM or coronary disease.

Disease definitions:

We used inclusive definitions of cardiomyopathy using the data described in this section, which will tend to provide upper estimates of the utility of opportunistic screening and surveillance. We use four approaches to identify CM:

1) Primary analyses use the "first occurrence" data

The first occurrence data provided the earliest date that an event or diagnosis of interest was reported for the UKBB participants. This data allowed for a time-specific assessment of the incidence of diagnoses; e.g. presence of a diagnosis after the date of recruitment up to the date of imaging.

2) Cardiac MRI imaging data

The diagnosis of participants with DCM via MRI imaging that do not have a previous diagnosis of DCM, allows for the identification of individuals that would be detected in a one-off assessment.

3) Hospital episode statistic data

The HES data is similar to 1) but it is not time-specific and allows for a DCM coded-specific analysis.

4) Non-ischaemic cardiomyopathy data

This inclusive definition of non-ischaemic cardiomyopathy allows for the identification of the maximum UKB participants that may present with non-ischaemic cardiomyopathy.

*Statistical analysis*

The association between variant carrier status and diagnoses were tested using Chi-squared or Fisher's exact tests. Lifetime survival analysis was performed with major adverse cardiovascular events (MACE) and death as the primary outcome and hazards ratios estimated and graphically presented as cumulative hazards curves using the *survival* and *survminer* packages in R. MACE consisted of heart failure (including cardiomyopathy that excluded HCM and coronary disease), stroke, atrial fibrillation or arrhythmia, and cardiac arrest.

*Calculation of the Number Needed to Treat (NNT)*

Although there was no significant difference in death based on TTNtv carrier status (P>0.05), the NNT was calculated based on the inverse of the adverse risk reduction of all-cause mortality (Table S1). The ARR was 0.9%, resulting in an NNT of 111.

*Estimation of background mortality*

Background mortality (Figure S2) was estimated as the mean death rate in 2019 of registered deaths in 59–69-year-olds in the UK by the Office for National Statistics (Reference number 12663). Counts of the number of UK male and female death registrations were summed for each single year of age, as were the mid-year population estimates for each single year of age. The death rate in 2019 for each year of age was calculated by dividing the counts of death for a single year of age by the corresponding population size. Mean death rate of 0.093% for 59-69-year-olds was calculated from the average of the resulting death rate values.

**Results**

*Variants identified*

The 38 variants in *FLNC* were predicted by VEP to cause the following consequences to the resulting protein: frameshift variant (n=14); splice acceptor variant (n=4); splice donor variant (n=4); stop gained (n=16).

The 487 variants in *TTN* were predicted by VEP to cause the following consequences to the resulting protein: frameshift variant (n=211); splice acceptor variant (n=15); splice donor variant (n=44); stop gained (n=209); stop gained, frameshift variant (n=8).

*Lifetime risk*

Lifetime risk of MACE was increased in *FLNC* heterozygotes (FLNCtv: n=10 (20%), no FLNCtv: n=21,802 (11%); HR=1.9, 95%CI=1.04-3.6, P=0.04), driven by increased risk of atrial fibrillation and arrhythmia (FLNCtv: n=9 (18%), no FLNCtv: n=16,294 (8%); HR=2.4, 95%CI=1.2-4.6, P=0.0096), with no significant difference in death or HF (P>0.05).

Lifetime risk of MACE was significantly increased in TTNtv heterozygotes (TTNtv: n=221 (25%), no TTNtv: n=21,591 (11%); HR=2.6, 95%CI=2.3-3.0, P=2.29x10$^{-46}$), driven by increased risk of atrial fibrillation and arrhythmia (TTNtv: n=171 (20%), no TTNtv: n=16,132 (8%); HR=2.7, 95%CI=2.3-3.1, P=1.67x10$^{-37}$), HF (TTNtv: n=99 (11%), no TTNtv: n=5,634 (3%); HR=4.4, 95%CI=3.6-5.3, P=5.37x10$^{-48}$), and CM (TTNtv: n=42 (5%), no TTNtv: n=663 (0.3%); HR=15.0, 95%CI=11.0-20.5, P=7.27x10$^{-65}$) with no significant difference in death, cardiac arrest, and stroke (P>0.05).

*Follow-up*

TTNtv heterozygotes (n=877) were followed up for 10,132 person-years, with a mean of 11.55 years per person. *FLNC* heterozygotes (n=50) were followed up for 569 person-years, with a mean of 11.39 years per person.

## Tables

**Table S1 a) Counts of CM and MACE phenotypes, and death, and estimates of CM prevalence, reported for the individuals of UKBB with exome sequencing data and a subset with imaging data available for analysis; b) Corresponding P-values of the burden analysis comparing heterozygotes to the rest of the population.**

Prevalence of disease was estimated using diagnostic codes at 3-time points; enrolment, date of imaging, and at the most recent assessment. This was completed to mimic the participants that would be identified by secondary findings; known affecteds at recruitment, unrecognised affecteds identified at imaging, and heterozygotes that developed disease during follow up. At the time of imaging, we assessed prevalence using diagnostic codes & imaging definition. Incident cases were identified between these time points. The association between variant carrier status and diagnoses were tested using Chi-squared (normal coloured cell) or Fisher's exact (pink cell) tests. Non-significant associations were highlighted in red font. *, for each trait the total needs subtraction of participants identified at previous incidence(s).

a) i)

Table S1a

| Timeline/Data | Description |
|---|---|
| **200,581 participants with whole exome sequencing data** | |
| Diagnostic code is present and first reported date precedes UKBB recruitment | Prevalence at ukbb recruitment |
| Diagnostic code is present and first reported date after UKBB recruitment | Incident cases post-recruitment (new cases) = value of surveillance |
| Diagnostic code is present (any date, "first reported date" combined dataset) | Cumulative prevalence |
| DCM diagnostic code is present (any date) using ICD code (HES data only) | Latest prevalence (these data allow for the most specific phenotype code for DCM, which is only available for HES data, not primary care or other data. Other estimates use non-ischaemic & non-hypertrophic CM as a surrogate) |
| Non-ischaemic cardiomyopathy (any date) | An upper-bound estimate based on an inclusive definition of non-ischaemic cardiomyopathies |
| **21,129 with whole exome sequencing data and cardiac MRI imaging** | |
| Diagnostic code is present and first reported date precedes UKBB recruitment | Prevalence at ukbb recruitment |
| Diagnostic code is present and first reported date after UKBB recruitment but precedes imaging | Incident cases between recruitment & imaging |
| Diagnostic code is present at time of CMR ("first reported date" combined dataset) | Prevalence of "known" DCM at time of imaging |
| Imaging criteria for DCM on CMR (increased LVEDV+ decreased LVEF without HCM or CAD), AND no CM reported before imaging date | Phenotype identified via MRI only |
| Diagnostic code present at time of CMR OR DCM criteria on CMR | Total DCM by diagnostic code or imaging |

a) ii)

| | | | TTNtv | | | | | | | | | | no TTNtv | | | | | |
|---|---|---|---|---|---|---|---|---|---|---|---|---|---|---|---|---|---|---|
| Proportion with DCM or CM (%) | n | DCM (n) | any CM (n) | cardiac arrest (n) | MACE (n) | AF/arrhythmia (n) | stroke (n) | heart failure (n) | death (n) | Prevalence of DCM or CM (%) | n | DCM (n) | any CM (n) | cardiac arrest (n) | MACE (n) | AF/arrhythmia (n) | stroke (n) | heart failure (n) | death (n) |
| 1.4 | 877 | | 12 | 1 | 80 | 59 | 13 | 24 | 0 | 0.09 | 199704 | | 172 | 82 | 7503 | 4634 | 2527 | 858 | 0 |
| 3.5 | 877* | | 30 | 5 | 141 | 112 | 6 | 75 | 60 | 0.25 | 199704* | | 491 | 894 | 14088 | 11498 | 953 | 4776 | 11859 |
| 4.8 | 877 | | 42 | 6 | 221 | 171 | 19 | 99 | 60 | 0.33 | 199704 | | 663 | 976 | 21591 | 16132 | 3480 | 5634 | 11859 |
| 4.3 | 877 | 38 | | | | | | | | 0.17 | 199704 | 338 | | | | | | | |
| 8.1 | 668 | 54 | | | | | | | | 0.73 | 158299 | 1149 | | | | | | | |
| 1.1 | 88 | | 1 | 0 | 10 | 8 | 1 | 2 | 0 | 0.02 | 21041 | | 5 | 3 | 572 | 406 | 148 | 31 | 0 |
| 2.3 | 88* | | 2 | 0 | 10 | 9 | 0 | 3 | 0 | 0.08 | 21041* | | 17 | 13 | 606 | 448 | 121 | 92 | 0 |
| 3.4 | 88 | | 3 | 0 | 20 | 17 | 1 | 5 | 0 | 0.10 | 21041 | | 22 | 16 | 1178 | 854 | 269 | 123 | 0 |
| 2.4 | 85 | 2 | | | | | | | | 0.38 | 21019 | 80 | | | | | | | |
| 5.7 | 88 | 5 | | | | | | | | 0.48 | 21041 | 102 | | | | | | | |
| | | | FLNCtv | | | | | | | | | | no FLNCtv | | | | | |
| Proportion with DCM or CM (%) | n | DCM (n) | any CM (n) | cardiac arrest (n) | MACE (n) | AF/arrhythmia (n) | stroke (n) | heart failure (n) | death (n) | | n | DCM (n) | any CM (n) | cardiac arrest (n) | MACE (n) | AF/arrhythmia (n) | stroke (n) | heart failure (n) | death (n) |
| 0 | 50 | | 0 | 0 | 5 | 5 | 0 | 0 | 0 | 0.09 | 200531 | | 184 | 83 | 7578 | 4688 | 2540 | 882 | 0 |
| 0 | 50* | | 0 | 0 | 5 | 4 | 0 | 3 | 5 | 0.26 | 200531* | | 521 | 899 | 14224 | 11606 | 959 | 4848 | 11914 |
| 0 | 50 | | 0 | 0 | 10 | 9 | 0 | 3 | 5 | 0.35 | 200531 | | 705 | 982 | 21802 | 16294 | 3499 | 5730 | 11914 |
| 0 | 50 | 0 | | | | | | | | 0.19 | 200531 | 376 | | | | | | | |
| 0 | 37 | 0 | | | | | | | | 0.76 | 158930 | 1203 | | | | | | | |
| 0 | 5 | | 0 | 0 | 0 | 0 | 0 | 0 | 0 | 0.03 | 21124 | | 6 | 3 | 582 | 414 | 149 | 33 | 0 |
| 0 | 5* | | 0 | 0 | 0 | 0 | 0 | 0 | 0 | 0.09 | 21124* | | 19 | 13 | 616 | 457 | 121 | 95 | 0 |
| 0 | 5 | | 0 | 0 | 0 | 0 | 0 | 0 | 0 | 0.12 | 21124 | | 25 | 16 | 1198 | 871 | 270 | 128 | 0 |
| 0 | 5 | 0 | | | | | | | | 0.39 | 21099 | 82 | | | | | | | |
| 0 | 5 | 0 | | | | | | | | 0.51 | 21124 | 107 | | | | | | | |

b)

| Table S1b | | | | | | | | | | |
|---|---|---|---|---|---|---|---|---|---|---|
| | | | | TTNtv heterozygotes compared to non-carriers | | | | | | |
| Timeline/Data | Description | DCM | any CM | cardiac arrest | MACE | AF/arrhythmia | stroke | heart failure | death | |
| **200,581 participants with whole exome sequencing data** | | | | | | | | | | |
| Diagnostic code is present and first reported date precedes UKBB recruitment | Prevalence at ukbb recruitment | | 4.97E-11 | 0.30 | 9.41E-17 | 6.99E-18 | 0.57 | 3.01E-12 | 1 | |
| Diagnostic code is present and first reported date after UKBB recruitment | Incident cases post-recruitment (new cases) = value of surveillance | | 6.96E-24 | 0.60 | 2.88E-25 | 7.02E-19 | 0.32 | 2.17E-32 | 0.26 | |
| Diagnostic code is present (any date, "first reported date" combined dataset) | Cumulative prevalence | | 1.73E-33 | 0.33 | 1.84E-42 | 4.91E-35 | 0.34 | 5.80E-51 | 0.26 | |
| DCM diagnostic code is present (any date) using ICD code (HES data only) | Latest prevalence (these data allow for the most specific phenotype code for D | 4.94E-39 | | | | | | | | |
| Non-ischaemic cardiomyopathy (any date) | An upper-bound estimate based on an inclusive definition of non-ischaemic ca | 2.72E-106 | | | | | | | | |
| **21,129 with whole exome sequencing data and cardiac MRI imaging** | | | | | | | | | | |
| Diagnostic code is present and first reported date precedes UKBB recruitment | Prevalence at ukbb recruitment | | 0.0247 | 1 | 0.0002 | 0.0003 | 0.46 | 0.0083 | 1 | |
| Diagnostic code is present and first reported date after UKBB recruitment but pre | Incident cases between recruitment & imaging | | 0.0028 | 1 | 0.0002 | 0.0001 | 1 | 0.0073 | 1 | |
| Diagnostic code is present at time of CMR ("first reported date" combined datase | Prevalence of "known" DCM at time of imaging | | 0.0002 | 1 | 7.70E-08 | 9.60E-08 | 1 | 0.0002 | 1 | |
| Imaging criteria for DCM on CMR (increased LVEDV+ decreased LVEF without | Phenotype identified via MRI only | 0.046 | | | | | | | | |
| Diagnostic code present at time of CMR OR DCM criteria on CMR | Total DCM by diagnostic code or imaging | 8.51E-05 | | | | | | | | |

| | | | | FLNCtv heterozygotes compared to non-carriers | | | | | | |
|---|---|---|---|---|---|---|---|---|---|---|
| Timeline/Data | Description | DCM | any CM | cardiac arrest | MACE | AF/arrhythmia | stroke | heart failure | death | |
| **200,581 participants with whole exome sequencing data** | | | | | | | | | | |
| Diagnostic code is present and first reported date precedes UKBB recruitment | Prevalence at ukbb recruitment | | 1 | 1 | 0.04 | 0.006 | 1 | 1 | 1 | |
| Diagnostic code is present and first reported date after UKBB recruitment | Incident cases post-recruitment (new cases) = value of surveillance | | 1 | 1 | 0.40 | 0.53 | 1 | 0.12 | 0.22 | |
| Diagnostic code is present (any date, "first reported date" combined dataset) | Cumulative prevalence | | 1 | 1 | 0.04 | 0.02 | 1 | 0.17 | 0.22 | |
| DCM diagnostic code is present (any date) using ICD code (HES data only) | Latest prevalence (these data allow for the most specific phenotype code for D | 1 | 1 | 1 | 0.09 | 0.19 | 1 | 0.16 | | |
| Non-ischaemic cardiomyopathy (any date) | An upper-bound estimate based on an inclusive definition of non-ischaemic ca | 1 | | | | | | | | |
| **21,129 with whole exome sequencing data and cardiac MRI imaging** | | | | | | | | | | |
| Diagnostic code is present and first reported date precedes UKBB recruitment | Prevalence at ukbb recruitment | | 1 | 1 | 1 | 1 | 1 | 1 | 1 | |
| Diagnostic code is present and first reported date after UKBB recruitment but pre | Incident cases between recruitment & imaging | | 1 | 1 | 1 | 1 | 1 | 1 | 1 | |
| Diagnostic code is present at time of CMR ("first reported date" combined datase | Prevalence of "known" DCM at time of imaging | | 1 | 1 | 1 | 1 | 1 | 1 | 1 | |
| Imaging criteria for DCM on CMR (increased LVEDV+ decreased LVEF without | Phenotype identified via MRI only | 1 | | | | | | | | |
| Diagnostic code present at time of CMR OR DCM criteria on CMR | Total DCM by diagnostic code or imaging | 1 | | | | | | | | |

Table S1 a) Counts of CM and MACE phenotypes, and death, and estimates of CM prevalence, reported for the individuals of UKBB with exome sequencing data and a subset with imaging data available for analysis; b) Corresponding P-values of the burden analysis comparing heterozygotes to the rest of the population.

Prevalence of disease was estimated using diagnostic codes at 3-time points; enrolment, date of imaging, and at the most recent assessment. This was completed to mimic the participants that would be identified by secondary findings; known affecteds at recruitment, unrecognised affecteds identified at imaging, and heterozygotes that developed disease during follow up. At the time of imaging, we assessed prevalence using diagnostic codes & imaging definition. Incident cases were identified between these time points. The association between variant carrier status and diagnoses were tested using Chi-squared (normal coloured cell) or Fisher's exact (pink cell) tests. Non-significant associations were highlighted in red font. *, for each trait the total needs subtraction of participants identified at previous incidence(s).

**Table S2 Prevalence of DCM and TTNtv in Published Population Cohorts**

The proportion of each cohort with TTNtv were between 0.4%-1.4% across three population cohorts. The prevalence of DCM in each cohort was in the range 0.05%-6%, consensus estimates are 0.4% in literature[8]. The proportion of the cohorts with DCM and TTNtv was 1.5%-30.3%. *cohort number differs in article text; presented is number from Table 1[9].

| Reference | Cohort | Participants (n) | TTNtv (PSI>90) in cohort (n) | DCM in cohort (n) | Proportion of TTNtv heterozygotes with DCM (n) |
|---|---|---|---|---|---|
| Haggerty et al. 2019[9] | PennMedicine BioBank | 10,289* | 142 (1.38%) | 613 (5.96%) | 43 (30.28%) |
| Haggerty et al. 2019[9] | Geisinger MyCode Community Health Initiative | 61,040 | 359 (0.59%) | 622 (1.02%) | 27 (7.52%) |
| Pirruccello et al. 2020[10] | UK Biobank baseline DCM | 49,944 | 227 (0.45%) | 26 (0.05%) | 4 (1.76%) |
| Pirruccello et al. 2020[10] | UK Biobank new DCM | 45,747 | 196 (0.43%) | 26 (0.06%) | 3 (1.53%) |

**Figures**

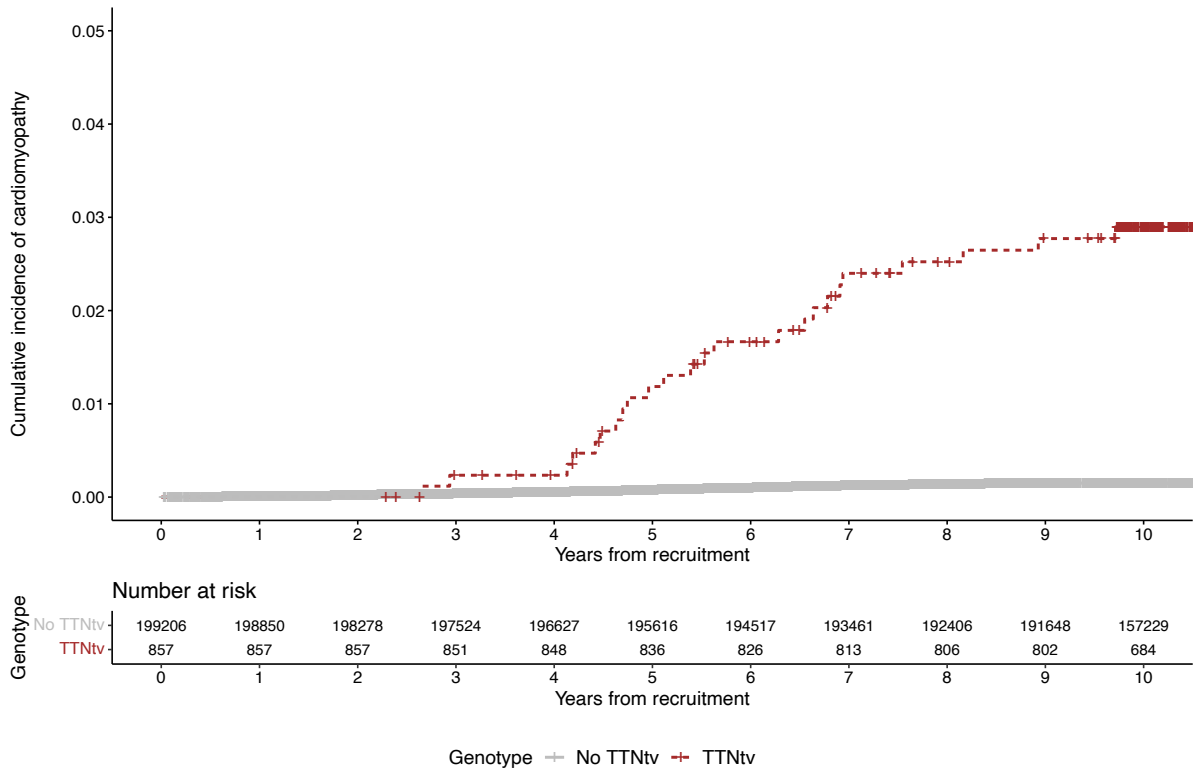

**Figure S1 Cumulative incidence curve of cardiomyopathy over ten years post recruitment to the UKBB, stratified by TTNtv carrier status**.

There were 2.8 events per 1,000 person-years. Individuals with cardiomyopathy and a code of HCM or coronary artery disease at any time (n=334; n=8 in TTNtv group) and individuals with cardiomyopathy at baseline (n=184; n=12 in TTNtv group) were excluded from the analysis.

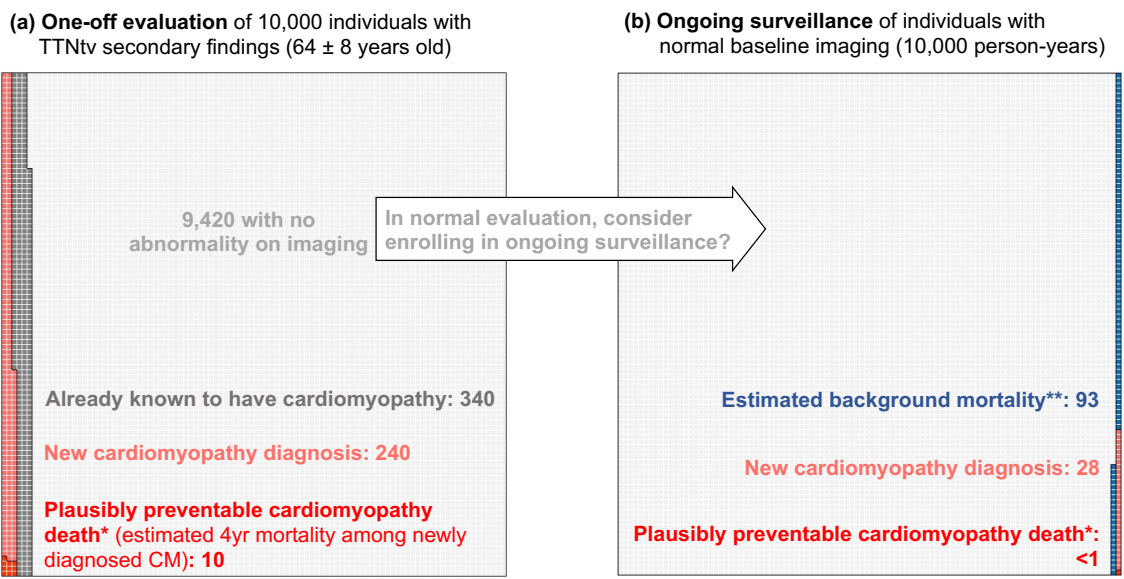

**Figure S2 Visual representation of the potential benefits of opportunistic screening for TTNtv estimated using UK Biobank data.**

A) Expected findings for a one-off cardiac imaging evaluation of 10,000 individuals with a TTNtv identified as a secondary finding. We estimate that 340 individuals would already be known to have cardiomyopathy, and 240 would be newly identified. CM-related mortality has been estimated as ~4%/4years[11,12], so we expect ~10 deaths in this time-frame amongst the newly identified cases, which might plausibly be preventable. This is one estimate of the benefit of opportunity screening, though it is not yet known to what extent early identification would prevent these, particularly given that this population will have other competing risks. Since they are undergoing clinical sequencing for another indication, the morbidity and mortality of that condition would influence the overall value of screening for secondary findings.

B) For the 9,420 individuals with no abnormality on initial imaging, we need to consider the potential value of ongoing surveillance, e.g., with serial imaging. For each 10,000 person-years of surveillance we expect 28 new cardiomyopathy diagnoses, with <1 plausibly preventable cardiomyopathy-related death in the next four years (based on mortality rates described above). For comparison, the estimated background mortality in the UK is 93 in 10,000 person-years. The background mortality in a population undergoing clinical sequencing for another indication may be higher.

These estimates are based on the UK biobank population, a population of older adults around the typical age of presentation for dilated cardiomyopathy. The yields of one-off and serial evaluation might be expected to be lower in younger individuals. The potential benefit of opportunistic screening for sudden cardiac arrest prevention can also be estimated by directly measuring the incidence of sudden cardiac arrest. Amongst TTNtv heterozygotes in the UK biobank (including individuals already known to have disease, as well as those newly recognised) the incidence is 5/10,132 person-years, i.e., 0.05%, compared with a background SCA incidence of 0.04% in the remainder of the cohort (P>0.05).

*Plausibly preventable cardiomyopathy mortality was estimated as 4%/4yr among newly diagnosed cardiomyopathy[11,12]. **mean death rate in 2019 of registered deaths in 59-69-year-olds in the UK by the Office for

National Statistics. The data supporting this figure can be found in the supplementary methods, Table S1, and Figure S1. Of note, these estimates do not account for competing risks.